\documentclass[twocolumn,showpacs,preprintnumbers,amsmath,amssymb]{revtex4}

\usepackage{graphicx}
\usepackage{dcolumn}
\usepackage{bm}

\begin{document}


\title{First-order phase transition of fixed connectivity surfaces}

\author{Hiroshi Koibuchi}
 \email{koibuchi@mech.ibaraki-ct.ac.jp}

\author{Nobuyuki Kusano}%
\author{Atsusi Nidaira}%
\author{Komei Suzuki}%
\affiliation{%
Department of Mechanical Engineering, Ibaraki College of Technology, Nakane 866 Hitachnaka, Ibaraki 312-8508, Japan}%

\author{Mitsuru Yamada}%

\affiliation{%
Department of Mathematical Sciences, Faculty of Sciences, Ibaraki University,\\ 
Bunkyo 2-1-1, Mito, Ibaraki 310-8512, Japan}%



\begin{abstract}
We report a numerical evidence of the discontinuous transition of a tethered membrane model which is defined within a framework of the membrane elasticity of Helfrich. Two kinds of phantom tethered membrane models are studied via the canonical Monte Carlo simulation on triangulated fixed connectivity surfaces of spherical topology. A surface model is defined by the Gaussian term and the bending energy term, and the other, which is tensionless, is defined by the bending energy term and a hard wall potential. The bending energy is defined by using the normal vector at each vertex. Both of the models undergo the first-order phase transition characterized by a gap of the bending energy. The phase structure of the models depends on the choice of discrete bending energy. 

\end{abstract}

\maketitle

\section{Introduction}\label{intro}
Tethered membrane models \cite{David-book,WHEATER-rev,Wiese-AP,Bowick-PR} are ordinarily defined by Hamiltonian that is a linear combination of discrete bending energy and surface tension energy \cite{HELFRICH,POLYAKOV}. Hence, there may be a variety of statistical models of membranes, since discrete Hamiltonian can be chosen arbitrarily even within Helfrich or Polyakov-Kleinert prescription of membranes. As a consequence, it is natural to ask whether the phase structure \cite{Peliti-Leibler,DavidGuitter,David,BKS,BK} of the model depends on the Hamiltonian.

However, little attention has been given to the dependence of the phase transition on the Hamiltonian of tethered surfaces both for models that have surface tension \cite{WHEATER,BCFTA,KY-2,KOIB-pla-2,CATTERALL,KY-1,KOIB-pla-1,AMBJORN,ABGFHHM,BCHHM} and for tensionless models \cite{KANTOR-NELSON,KANTOR,KANTOR-KARDAR-NELSON,Ho-Baum,Gompper-Kroll,Bowick-EPJ5E}. Almost all numerical studies done so far utilize the bending energy of the ordinary form $1\!-\!{\bf n}_i\cdot{\bf n}_j$, where ${\bf n}_i$ is the normal vector of the triangle $i$. 

One other discrete bending energy that has been utilized by Gompper et.al \cite{Gompper} is based on the discretization of the Laplacian in the dual lattice formulation of discrete mechanics by T.D.Lee \cite{TDLEE}. Similar discrete bending energy was adopted in Refs. \cite{KY-2,KY-1,KY-0}. Both discrete bending energies give results compatible with the continuous phase transition of the model \cite{Peliti-Leibler,DavidGuitter,David,BKS,BK}.

 Recently, it was reported \cite{KD} that a tethered membrane model with the ordinary bending energy undergoes the discontinuous phase transition predicted in \cite{PKN}, although the L-J potential is assumed to serve as the Gaussian term. Hence, we think it is worthwhile to show that the discontinuous phase transition can be seen in a tethered membrane model when the Hamiltonian is defined only by a discretization of Helfrich Hamiltonian.  

The purpose of this paper is to show numerical evidence that the phase structure of phantom tethered models depends on the choice of the discrete bending energy. By using the normal vector at each vertex, we define a bending energy which is different from the ordinary one. We will study two kinds of models; one is a model that has the Gaussian term for surface tension and the other is a tensionless model that has no surface tension term but has a hard wall potential. It will be shown that both models undergo the first-order phase transition. 

\section{Model and MC techniques}
Membrane models are ordinarily defined by the discrete Hamiltonian $S\!=\!S_1\!+\!bS_2$ with the bending rigidity $b$, where $S_1$ is the surface tension energy and $S_2$ is the bending energy respectively defined by 
\begin{equation}
\label{Disc-Eneg-1} 
S_1=\sum_{(ij)} \left(X_i-X_j\right)^2,\quad S_2=\sum_{(ij)}\left(1-{\bf n}_i\cdot {\bf n}_j\right).
\end{equation} 
$\sum_{(ij)}$ in Eq.(\ref{Disc-Eneg-1}) is over all bonds $(ij)$, and ${\bf n}_i$, ${\bf n}_j$ are the unit normal vectors of the triangles sharing the bond  $(ij)$. $X_i(\in {\bf R}^3)$ in $S_1$ is the position of the vertex $i$.

Other possible bending energy $S_2$ can be obtained by using the normal vector of the vertex $i$ such as
\begin{equation}
\label{normal}
{\bf n}(i)= {{\bf N}_i \over \vert {\bf N}_i\vert}, \quad {\bf N}_i = \sum _{j(i)} {\bf  n}_{j(i)} A_{\it \Delta_{j(i)}},
\end{equation}
where $\sum _{j(i)}$ denotes the summation over triangles  $j(i)$ linked to the vertex $i$. The vector ${\bf n}_{j(i)}$ is the unit normal of the triangle $j(i)$, and  $A_{\it \Delta_{j(i)}}$ is the area of $j(i)$. 

 The new discrete bending energy can be obtained by using the normal vector of Eq.(\ref{normal}). Thus, we have
\begin{equation}
\label{Disc-Eneg-2} 
 S_2=\sum_i\sum_{j(i)}\left[1-{\bf n}(i)\cdot {\bf n}_{j(i)}\right],
\end{equation} 
which is clearly different from that of Eq.(\ref{Disc-Eneg-1}). It should be noted that $S_2({\rm illdef})=\sum_{i,j}\left(1-{\bf n}(i)\cdot {\bf n}(j)\right)$ defined only by using the normal vectors ${\bf n}(i)$ in Eq.(\ref{normal}) is not well defined. This ill-definedness comes from the fact that there exists two surfaces locally different from each other that has the same value of $S_2({\rm illdef})$. Two normal vectors at the ends of a bond $(i,j)$ can be parallel for surfaces that are not smooth.

We study two kinds of models in this paper. The first, which will be denoted by Model-1, is a model defined by
\begin{eqnarray} 
\label{Part-Func-1}
 Z_1 =  \int \prod _{i=1}^{N} d X_i \exp[-(S_1 + b S_2)],\qquad ({\rm Model\!-\!1})\nonumber \\
  S_1=\sum_{(ij)} \left(X_i-X_j\right)^2,\; S_2=\sum_i\sum_{j(i)}\left(1-{\bf n}(i)\cdot {\bf n}_{j(i)}\right),
\end{eqnarray} 
where the center of the surface is fixed to remove the translational zero mode.  $S_2$ is identical with (\ref{Disc-Eneg-2}).

The second model, which will be denoted by Model-2, is a tensionless model defined by
\begin{eqnarray} 
\label{Part-Func-2}
 Z_2 =  \int \prod _{i=1}^{N} d X_i \exp[-(b S_2+V)],\qquad ({\rm Model\!-\!2})\nonumber \\
 S_2=\sum_i\sum_{j(i)}\left(1-{\bf n}(i)\cdot {\bf n}_{j(i)}\right),
\end{eqnarray} 
where $S_2$ is identical with that of Model-1 in Eq.(\ref{Part-Func-1}), and $V$ is the hard wall potential defined by
\begin{equation}
\label{V}
V(|X_i-X_j|)= \left\{
       \begin{array}{@{\,}ll}
       0 & \quad (0< |X_i-X_j| < r_0), \\
      \infty & \quad ({\rm otherwise}). 
       \end{array}
       \right. 
\end{equation}
The value of $r_0$ in the right hand side of (\ref{V}) is fixed to $r_0\!=\!\sqrt{1.15}$. As a consequence we have $\langle \sum (X_i\!-\!X_j)^2 \rangle /N \simeq 3/2$, which holds for Model-1 which contains the Gaussian term $S_1$. It should be noted that Model-2 is considered to be independent of the hidden length introduced by $r_0$. The MC results are independent of the value of $r_0$. This was, in fact, precisely checked in Ref. \cite{KOIB-pla-2}.

\begin{figure}[bht]
\centering
\includegraphics[width=7cm]{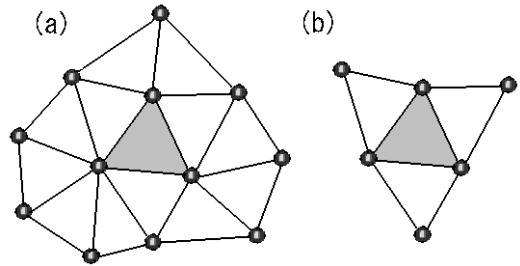}
 \caption{(a) Ranges of interaction between normal vectors of triangles for $S_2$ in (\ref{Disc-Eneg-2} ), and (b) those for $S_2$ in (\ref{Disc-Eneg-1}). The normal vector of the shaded triangle interacts with those of the surrounding triangles in (a) and (b). Small spheres represent vertices. }
 \label{fig-1}
\end{figure}
Figures \ref{fig-1} (a) and \ref{fig-1} (b) show the range of interactions described by $S_2$ in (\ref{Disc-Eneg-2}) and the ordinary $S_2$ in (\ref{Disc-Eneg-1}). A difference between $S_2$ in (\ref{Disc-Eneg-2}) and $S_2$ in (\ref{Disc-Eneg-1}) can be seen in the number of triangles whose normal vectors interact with the one of a given triangle, which is shaded in Figs. \ref{fig-1} (a) and \ref{fig-1} (b). The number of triangles for $S_2$ in (\ref{Disc-Eneg-2}) is dependent on a given triangle and hence locally changes, while the number for $S_2$ in (\ref{Disc-Eneg-1}) is always 3. 

We use the canonical Metropolis Monte Carlo technique. Spheres are triangulated by linking uniformly scattered points. The histograms of coordination number of surfaces are identical with those shown in Ref. \cite{KOIB-pla-2}. 

The position $X$ of vertices is updated in MC by moving the current position $X$ to a new position $X^\prime \!=\!X\!+\!\delta X$, where $\delta X$ is chosen in a small sphere by using uniform random numbers. The radius $R_0$ of the small sphere is fixed to $R_0\!=\! \epsilon l_0$, where $l_0$ is the mean bond length which is computed at every 250 MCS (Monte Carlo Sweeps), and a constant $\epsilon$ is fixed at the beginning of the simulation to maintain $50\sim 55 \%$ acceptance rate for Model-1 and $55\sim 65 \%$ for Model-2. The radius $R_0$ becomes almost constant, because $l_0$ is constant in the equilibrium configurations. 

We impose the lower bound $10^{-6} A_0$ on the area of triangles, where $A_0$ is the mean area of triangles computed at every 250 MCS. As a consequence, the update of $X$ are constrained so that the resulting area of triangles becomes larger than $10^{-6} A_0$. However, areas of almost all triangles are larger than $10^{-6} A_0$ in our MC without the lower bound, hence it seems that the areas are almost free from such constraint. No constraint is imposed on the bond length.

\section{Results}
\begin{figure}[bht] 
\centering
\includegraphics[width=8.5cm]{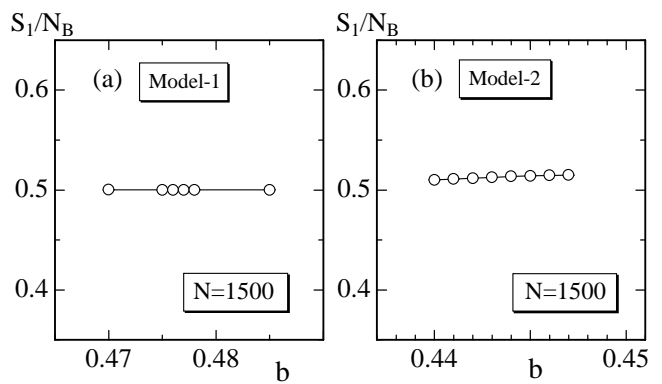}
 \caption{(a) $S_1/N_B$ vs  $b$ of Model-1, and (b) $S_1/N_B$ vs  $b$ of Model-2, where $N_B$ is the total number of bond. $N\!=\!1500$. }
 \label{fig-2}
\end{figure} 
We first show $S_1/N_B$ of Model-1 and Model-2 respectively in Figs. \ref{fig-2} (a) and \ref{fig-2} (b), where $N_B$ is the total number of bond. It should be noted that $S_1/N_B$ is the mean bond length squares $l_0^2$. $S_1/N_B$ in  Fig. \ref{fig-2} (a) of Model-1 is completely compatible with the expected result $S_1/N\!=\!3/2$, since $N_B\!=\!3N\!-\!6 (\simeq 3N)$ on the spherical surfaces. In fact, a typical sample in  Fig. \ref{fig-2} (a) is $S_1/N_B\!=\! 0.50015\!\pm\!0.00012$ at $b\!=\! 0.476$. Moreover, $S_1/N_B$ in  Fig. \ref{fig-2} (b) of Model-2 is also compatible with our expectation $S_1/N\!\simeq\!3/2$ as already stated in the paragraph below Eq. (\ref{V}), although the Gaussian term $S_1$ is not included in the Hamiltonian of the Model-2. Thus, we confirmed that $l_0$ is constant in the equilibrium configurations in both models.

The specific heat $C_{S_2}$ is a fluctuation of the bending energy and is given by
\begin{equation}
\label{Spec-Heat}
C_{S_2} = {b^2\over N} \left( \langle S_2^2 \rangle - \langle S_2 \rangle ^2 \right).
\end{equation}
\begin{figure}[bht] 
\centering
\includegraphics[width=8.5cm]{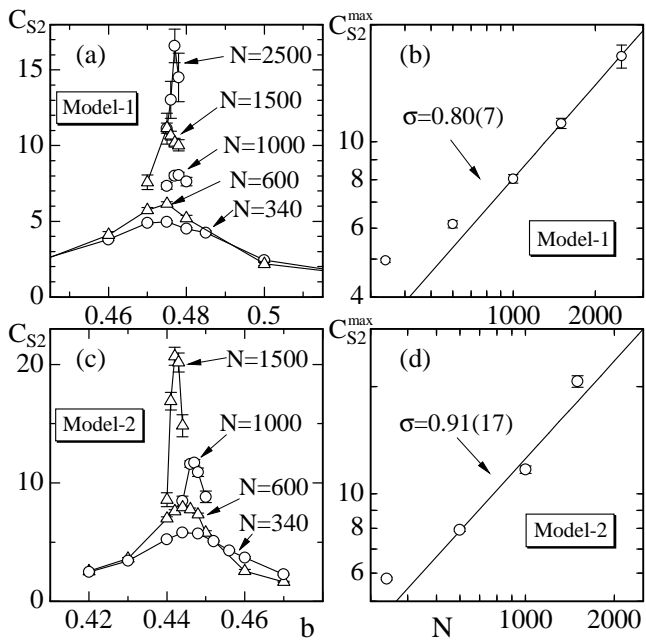}
 \caption{(a) $C_{S_2}$ vs  $b$, and (b) $C_{S_2}^{\rm max}$ vs  $N$ in log-log scale, both (a) and (b) are obtained by Model-1 whose Hamiltonian is $S_1\!+\!bS_2$. (c) $C_{S_2}$ vs  $b$, and (d) $C_{S_2}^{\rm max}$ vs  $N$ in log-log scale, both (c) and (d) are obtained by Model-2 whose Hamiltonian is $bS_2\!+\!V$. }
 \label{fig-3}
\end{figure} 
Total number of MCS is about $0.8\!\sim\!1.0\!\times\! 10^8$ for $N\!=\!340, N\!=\!600$, $1.5\!\times\! 10^8$ for $N\!=\!1000$, $3\!\times\! 10^8$ for $N\!=\!1500$, and $2.2\!\times\! 10^8$ for $N\!=\!2500$ at the transition points $b_c(N)$ for Model-1. Number of MCS at $b\!\not=\!b_c(N)$ is relatively small. Total number of MCS for Model-2 is smaller than that for Model-1, since the speed of convergence of Model-2 is relatively faster than Model-1.

Figure \ref{fig-3} (a)  shows $C_{S_2}$ vs $b$ of Model-1. The peak values $C_{S_2}^{\rm max}$ of Model-1 are plotted in Fig. \ref{fig-3} (b) against $N$ in log-log scale. Figures \ref{fig-3} (c) and \ref{fig-3} (d) are results obtained by Model-2.  The number of molecules are $N\!=\!340$, $N\!=\!600$, $N\!=\!1000$, $N\!=\!1500$ for Model-2. 

 The slope of the straight lines in Fig. \ref{fig-3} (b) and Fig. \ref{fig-3} (d)  represents the critical exponent $\sigma$ defined by
\begin{equation}
\label{Def-sigma}
C_{S_2}^{\rm max} \sim N^\sigma.
\end{equation}
The largest three data in each figure are included in the fit, and we have
\begin{eqnarray}
\label{Result-sigma}
\sigma_1 = 0.798(66)  \qquad &&({\rm Model\!-\!1}), \nonumber \\
\sigma_2=0.914(166) \qquad &&({\rm Model\!-\!2}).   
\end{eqnarray}
The value $\sigma_1 \!=\! 0.798(66)$ is smaller than 1 and hence implies that the order of the phase transition of Model-1 is of second-order. However, as we will see next, the order of the phase transition of Model-1 is considered to be of first order.  While $\sigma_1 \!<\! 1$, the value $\sigma_2 \!=\! 0.914(166)$ almost equals to 1 and hence suggests that Model-2 undergoes the first-order phase transition.

\begin{figure}[bht]
\centering
\includegraphics[width=8.5cm]{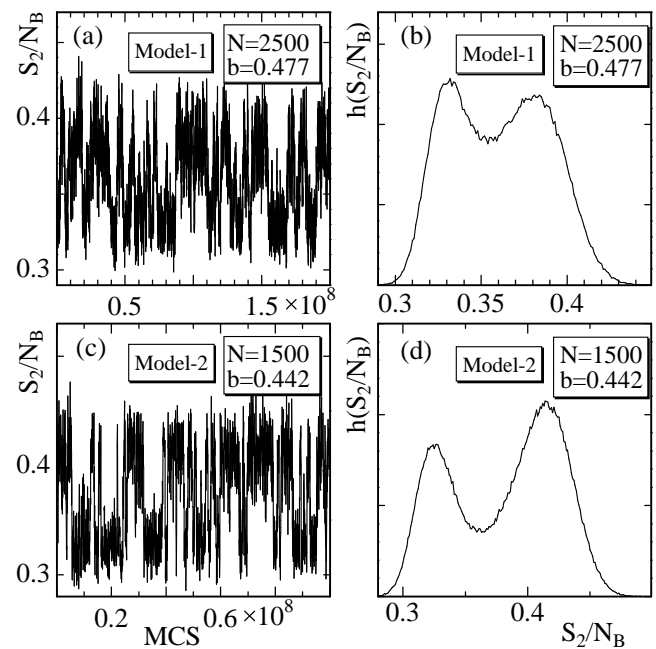}
 \caption{(a) Variation of $S_2/N_B$ against the number of MCS, and (b) the histogram $h(S_2/N_B)$, obtained by Model-1. The results obtained by Model-2 are shown in (c) and (d).}
 \label{fig-4}
\end{figure}
To clarify the order of the transition of Model-1, we plot in Fig. \ref{fig-4} (a) the variation of $S_2/N_B$ against the number of MCS. The series $S_2$ shown in Fig. \ref{fig-4} (a) was obtained at every $5\times 10^4$ MCS at the transition point $b\!=\!b_c(N)$ on the surface of size $N\!=\!2500$. The corresponding histogram $h(S_2/N_B)$ is drawn in Fig. \ref{fig-4} (b). Figures \ref{fig-4} (c) and \ref{fig-4} (d) are the results obtained by Model-2 of size $N\!=\!1500$.

We clearly see in Fig. \ref{fig-4} (a) that there are two distinct states which represent a discontinuous phase transition in Model-1. The histogram in Fig. \ref{fig-4} (b) shows more clearly the existence of the two states separated by a gap of $S_2$ in Model-1. It is also easy to understand from Figs. \ref{fig-4} (c) and \ref{fig-4} (d) that Model-2 undergoes the first-order phase transition characterized by a gap of $S_2$. 

\begin{figure}[bht]
\centering
\includegraphics[width=8.5cm]{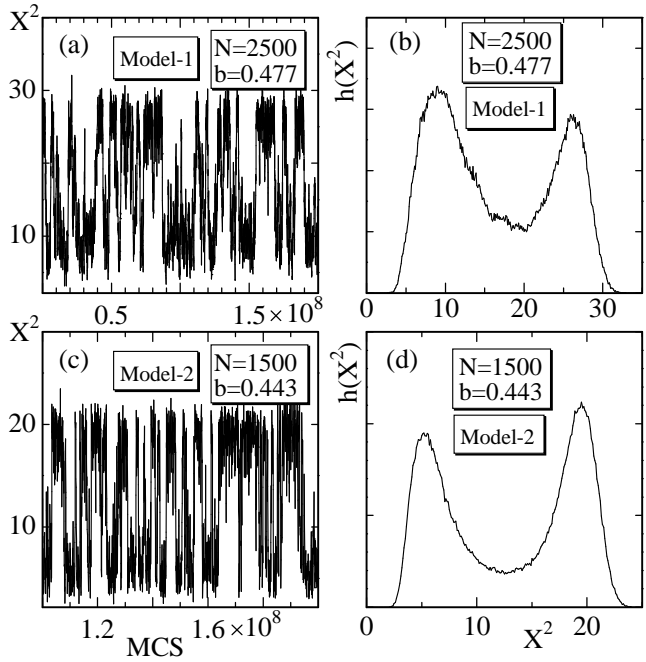}
 \caption{(a) Variation of $X^2$ against the number of MCS, and (b) the histogram $h(X^2)$,  obtained by Model-1. The results obtained by Model-2 are shown in (c) and (d).}
 \label{fig-5}
\end{figure}
The mean square size $X^2$ defined by
\begin{equation}
\label{X2}
X^2={1\over N} \sum_i \left( X_i -\bar X \right)^2,\qquad \bar X ={1\over N} \sum_i X_i,
\end{equation}
is plotted in Fig. \ref{fig-5}(a) against the number of MCS of Model-1.  The corresponding histogram $h(X^2)$ is drawn in Fig. \ref{fig-5}(b). Figures \ref{fig-5} (c) and \ref{fig-5} (d) are the results obtained by Model-2. We see two different sizes at the transition point in each model, hence consider that the phase transitions in both models are characterized also by discontinuity of $X^2$. The reason why we use $X^2$ obtained at $b\!=\!0.443$ in Figs. \ref{fig-5} (c),(d) is that the double peaks in the histogram of $X^2$ at $b\!=\!0.443$ is clearer than at $b\!=\!0.442$ where the histogram of $S_2/N_B$ plotted in Figs. \ref{fig-4} (c),(d) were obtained. 

The Hausdorff dimension \cite{Gross,Duplantier,JK} is defined by
\begin{equation}
\label{Hausdorff}
X^2\sim N^{2/H}.
\end{equation}
The gap of $X^2$ at the transition point implies that $H$ discontinuously changes at that point. 

\begin{figure}[bht]
\centering
\includegraphics[width=8.5cm]{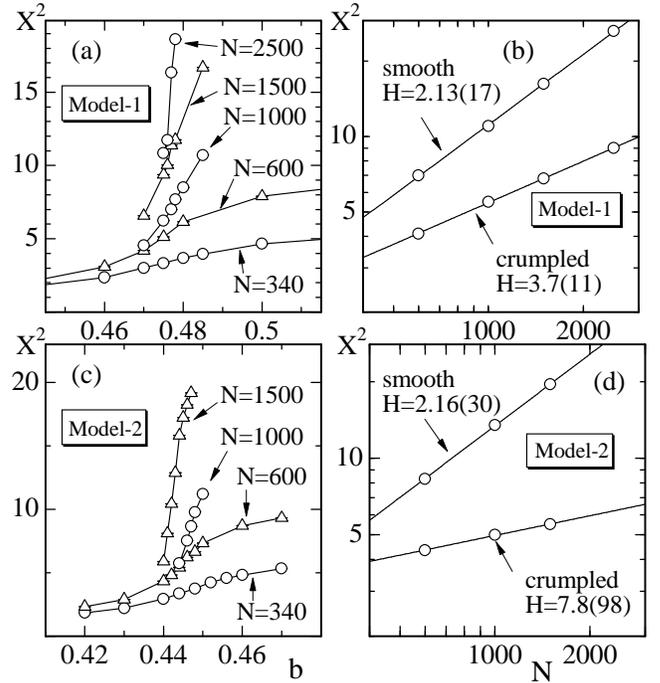}
 \caption{(a) $X^2$ vs $b$ of Model-1, (b) $X^2$ vs $N$ at the transition point $b\!=\!b_c(N)$. The results obtained by Model-2 are shown in (c) and (d).}
 \label{fig-6}
\end{figure}
We plot in Fig. \ref{fig-6} (a) $X^2$ vs $b$ of Model-1. The mean square size $X^2$ at $b\!=\!b_c(N)$ are plotted against $N$ in log-log scale in Fig. \ref{fig-6}(b). The straight line denoted by {\it smooth} is obtained by fitting $X^2$, each of which is the larger $X^2$ in the double peaks shown in Fig. \ref{fig-5} (b). Another straight line denoted by {\it crumpled} is obtained by fitting the smaller $X^2$  in the peaks. Errors of $X^2$ were not included in the least squares fitting, since the fitting was done by using only the peak values of $X^2$ in the histogram shown in Fig. \ref{fig-5} (b). Figures \ref{fig-6} (c) and \ref{fig-6} (d) show the results of Model-2. 

From the slope of the straight lines in Figs. \ref{fig-6} (b) and \ref{fig-6} (d), we have 
\begin{eqnarray}
\label{Result-H}
H_1(\uparrow)=2.13(17),\; H_1(\downarrow)=3.66(107)  &&({\rm Model\!-\!1}), \nonumber \\
H_2(\uparrow)=2.16(30),\; H_2(\downarrow)=7.84(977)  &&({\rm Model\!-\!2}).  \nonumber \\
\end{eqnarray}
$H(\uparrow) (H(\downarrow))$ is considered as the Hausdorff dimension in the smooth (crumpled) phase at $b\!>\!b_c$ ($b\!<\!b_c$) just above (below) $b_c$ in each model. The reason of the large errors both in $H_1(\downarrow)$ and in $H_2(\downarrow)$ seems come from the fact that there is a few data points of $X^2$ included in the fitting.

We understand from the straight lines in Figs. \ref{fig-6} (b) and \ref{fig-6} (d) that the phase transition of Model-2 is relatively stronger than that of Model-1, although both of the transition are the first order. The gap of $H$ at $b\!=\!b_c$ of Model-2 is relatively larger than that of Model-1; this difference of $H$ can be visible in the slope of the straight lines in Figs. \ref{fig-6} (b) and \ref{fig-6} (d). 

There is no difference between the surfaces in the smooth phase of the models in this paper and those of \cite{KOIB-pla-2}. While the surfaces in the disordered (or crumpled) phase of the models in this paper are more crumpled than those in \cite{KOIB-pla-2}. The Hausdorff dimension at $b\!>\!b_c$ of the models and those of \cite{KOIB-pla-2} are comparable, although the order of the transition of the models in this paper is different from that in \cite{KOIB-pla-2}; the models in \cite{KOIB-pla-2} have the continuous phase transition. 

We note also that both $H_1(\uparrow)$ and $H_2(\uparrow)$ are compatible with ( or slightly smaller than) the Flory prediction $H\!=\!2.5$, and they are almost compatible with an analytical result $H=2.39(23)$ which corresponds to the scaling exponent $\nu\!=\!0.84\!\pm\!0.04$ in \cite{David-Wiese-PRL96} where $\nu\!=\!2/H$. The values $H_1(\uparrow)$ and $H_2(\uparrow)$ in Eq. (\ref{Result-H}) imply that the surfaces are relatively swollen and smooth in the smooth phase at $b\!>\!b_c$ in both models.

\begin{figure}[bht]
\centering
\includegraphics[width=8.5cm]{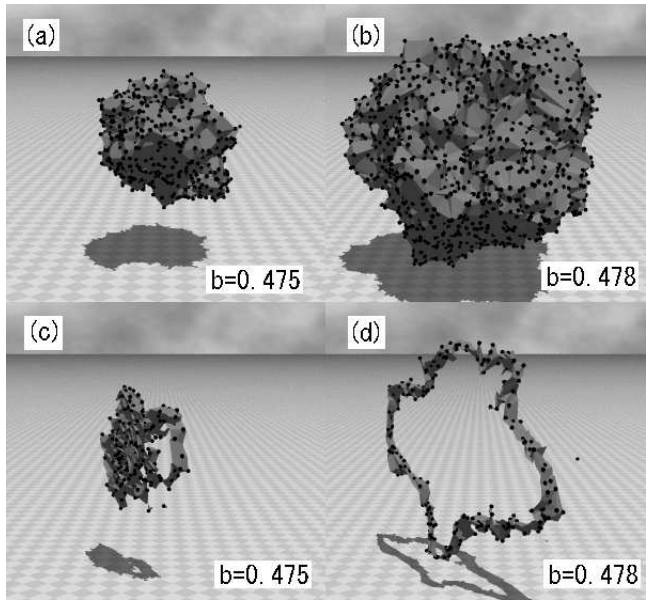}
 \caption{Snapshots of Model-1 surfaces  obtained at (a) $b\!=\!0.475$(crumpled phase), (b) $b\!=\!0.478$(smooth phase), and the sections of the surfaces in (a) and (b) are shown in (c) and (d) respectively. $N\!=\!2500$.}
 \label{fig-7}
\end{figure}
In order to see the surfaces, we show snapshots of size $N\!=\!2500$ of Model-1 in Figs. \ref{fig-7} (a) and \ref{fig-7} (b); one of which is obtained in the crumpled phase at $b\!=\!0.475$ and the other in the smooth phase at $b\!=\!0.478$. The sections of them are shown in Figs. \ref{fig-7} (c), (d). The surface swells in the smooth phase as expected. We also find that the surface in  Fig. \ref{fig-7} (b) is smooth only at long range scales and rough at short scales. This is compatible with that seen in the model with the ordinary bending energy \cite{KOIB-pla-2}. The surfaces of Model-2 are almost the same  as those in Fig. \ref{fig-7}.

\section{Summary and conclusions}
We have shown that the continuous phase transition seen in the ordinary tethered membrane models is strengthened in two kinds of tethered membrane models, whose bending energy is defined by using the normal vectors at the vertices. One of the models is defined by the Hamiltonian $S_1\!+\!bS_2$,  and the other is a tensionless model defined by $bS_2\!+\!V$, where $V$ is a hard wall potential. It was shown by extensive MC simulations that both of the models undergo the first-order phase transition which is characterized by a gap of $S_2$. The size of spherical surfaces and the Hausdorff dimension discontinuously change at the phase transition in both models. 

The definition of the Hamiltonian remains in the framework of membrane elasticity of Helfrich. The bending energy in Eq. (\ref{Disc-Eneg-2}) utilized in this paper appears to induce a non-nearest neighbor interaction between normal vectors of the surface. In fact, the range of the interaction is a bit larger than that of the ordinary bending energy as depicted in Fig. \ref{fig-1}. However, the bending energy in Eq. (\ref{Disc-Eneg-2}) is written by the normal vectors of Eq. (\ref{normal}) and the normal vectors of the neighboring   triangles, and hence it is defined only by local geometric quantities of the surface just like the ordinary bending energy in Eq. (\ref{Disc-Eneg-1}).

{\bf Acknowledgment}

This work is partially supported by the Grant-in-Aid for Scientific Research 15560160. 



\end{document}